\documentclass[twocolumn]{webofc}

\usepackage[varg]{txfonts}   
\usepackage{graphicx}
\usepackage{amsmath}
\usepackage{amssymb}
\usepackage{hyperref}
\usepackage{braket}
\usepackage[percent]{overpic}  

\newcommand{\ba}{\begin{eqnarray}}
\newcommand{\ea}{\end{eqnarray}}
\newcommand{\bsub}{\begin{subequations}}
\newcommand{\esub}{\end{subequations}}

\def\ket#1{|#1\rangle}

\begin{document}
%
\title{Intertwined quantum phase transitions in the Zr chain}
%
%

\author{\firstname{N.} \lastname{Gavrielov}\inst{1}\fnsep
\thanks{\email{noam.gavrielov@mail.huji.ac.il}} \and
\firstname{A.} \lastname{Leviatan}\inst{1}\fnsep
\thanks{\email{ami@phys.huji.ac.il}} \and
\firstname{F.} \lastname{Iachello}\inst{2}\fnsep 
\thanks{\email{francesco.iachello@yale.edu}}
}

\institute{Racah Institute of Physics, The Hebrew University, 
Jerusalem 91904, Israel
\and
Center for Theoretical Physics, Sloane Physics Laboratory, 
Yale University, New Haven, Connecticut 06520-8120, USA
          }

\abstract{
We introduce the notion of intertwined quantum phase transitions
(IQPTs), for which a crossing of two configurations
coexists with a pronounced shape-evolution of each configuration.
A detailed analysis in the framework of the interacting boson model
with configuration mixing, provides evidence for this scenario in
the Zr isotopes. The latter exhibit a normal configuration which remains
spherical along the chain, but exchanges roles with an
intruder configuration, which undergoes first a spherical to
prolate-deformed [U(5)$\to$SU(3)] QPT and then a crossover to
$\gamma$-unstable [SU(3)$\to$SO(6)].
}
\maketitle
Quantum phase transitions (QPTs) are qualitative changes in the properties
of a physical system, induced by a variation of parameters that appear in
the quantum Hamiltonian.
In nuclear physics, most of the attention in the study of QPTs,
has been devoted to shape phase transitions in a single
configuration, described by a single Hamiltonian,
$\hat{H} = \left( 1-\xi \right) \hat{H}_{1}+\xi \hat{H}_{2}$, 
where $\xi $ is the control parameter. As $\xi $ changes from $0$ to $1$,
the eigenvalues of the system change from those of 
$\hat{H}_{1}$ to those of $\hat{H}_{2} $. For sake of clarity, 
we denote these phase transitions Type~I. The latter have been observed
in the neutron number 90 region, {\it e.g.}, for Nd-Sm-Gd isotopes
~\cite{cejnar}.

A different type of phase transitions occurs when two (or more)
configurations coexist~\cite{Heyde11}.
In this case, the quantum Hamiltonian has a matrix form
with entries: $\hat{H}_{A}(\xi^A),\,\hat{H}_{B}(\xi^B),\,\hat{W}(\omega)$, 
where the indices $A$, $B$ denote the two configurations
and $\hat{W}$ denotes their coupling. We call for sake of
clarity these phase transitions Type~II~\cite{frank},
to distinguish them from those
of a single configuration.
Type~II QPTs have been observed in nuclei near shell closure,
{\it e.g.}, in the light Pb-Hg isotopes~\cite{ramos14},
albeit with strong mixing between the two configurations.
In the present contribution, we explore a situation where
in parallel to the crossing, each configuration maintains its purity and its
own shape-evolution with nucleon number.
We refer to such a scenario as intertwined quantum phase transitions (IQPTs)
in the sense that Type~I and Type~II coexist, and show empirical evidence for it
in the Zr chain~\cite{GavLevIac19}.

The $_{40}$Zr isotopes have been recently the subject of several
experimental~\cite{pietralla,Ansari17,Paul17,Witt18,Singh18}
and theoretical investigations,
including mean-field based methods~\cite{delaroche,mei,nomura16}
and the Monte-Carlo shell-model (MCSM)~\cite{taka}.
We adapt here the algebraic approach of the
Interacting Boson Model (IBM)~\cite{ibm},
with bosons representing valence nucleon pairs counted from the
nearest closed shells. This provides a simple tractable framework,
where phases of quadrupole shapes: spherical, prolate-deformed and
$\gamma$-unstable correspond to U(5), SU(3) and SO(6)
dynamical symmetries.

To be specific, we use the configuration mixing model
(IBM-CM) of~\cite{duval}, and
write the Hamiltonian not in matrix form, but rather in the equivalent form
\ba
\hat{H}=\hat{H}_{A}^{(N)}+\hat{H}_{B}^{(N+2)}+\hat{W}^{(N,N+2)} ~,
\label{Hamilt}
\ea
where $\hat{\cal O}^{(N)}=\hat{P}_{N}^{\dag }\hat{\cal O}\hat{P}_{N}$ 
and $\hat{\cal O}^{(N,N^{\prime })}=
\hat{P}_{N}^{\dag }\hat{\cal O}\hat{P}_{N^{\prime }}$, 
for an operator $\hat{\cal O}$, with $\hat{P}_{N}$, a projection operator 
onto the $[N] $ boson space. Here 
$\hat{H}_{A}^{(N)}$ represents the so-called normal 
($N$ boson space) configuration and $\hat{H}_{B}^{(N+2)}$
represents the so-called intruder ($N\!+\!2$ boson space)
configuration, which we have assumed, 
as in~\cite{sambataro} where a similar calculation was done for the 
$_{42}$Mo isotopes, to be a proton excitation across the subshell closure 
at proton number 40 (see Fig.~1 of~\cite{sambataro}). 
The explicit form of these Hamiltonians is 
\begin{subequations}
\label{Hmat}
\begin{align}
\hat{H}_{A} =&\,\epsilon _{d}^{(A)}\hat{n}_{d}
+\kappa ^{(A)}\hat{Q}_{\chi}\cdot \hat{Q}_{\chi } ~,
\label{HA}
\\
\hat{H}_{B} =&\,\epsilon _{d}^{(B)}\hat{n}_{d} 
+\kappa ^{(B)}\hat{Q}_{\chi}\cdot \hat{Q}_{\chi} 
+\kappa ^{\prime (B)}\hat{L}\cdot \hat{L} 
+ \Delta_p ~, 
\label{HB}
\\
\hat{W} =&\,\omega\,[\,( d^{\dag }\times d^{\dag })^{(0)}
+ \,(s^{\dag })^2\,] + {\rm H.c.} ~,
\label{W}
\end{align}
\end{subequations}
where the quadrupole operator is defined as 
$\hat{Q}_{\chi} = d^{\dag}s+s^{\dag }\tilde{d}
+\chi ( d^{\dag}\times \tilde{d}) ^{(2)}$ and
$\hat{n}_d$ is the $d$-boson number operator.
In Eq.~(\ref{HB}), $\Delta_p$ is the off-set 
between the normal and intruder configurations, where the index $p$ denotes
the fact that this is a proton excitation.
The resulting eigenstates $\ket{\Psi;L}$ 
with angular momentum $L$, are linear combinations of the
wave functions, $\Psi_A$ and $\Psi_B$,
in the two spaces $[N]$ and $[N+2]$,
\ba
\label{wf}
\ket{\Psi; L} = a\,\ket{\Psi_A; [N], L} + b\,\ket{\Psi_B; [N+2], L} ~,
\ea
with $a^{2}+b^{2}=1$.
\begin{figure}[t]
\centering
\includegraphics[width=1\linewidth]{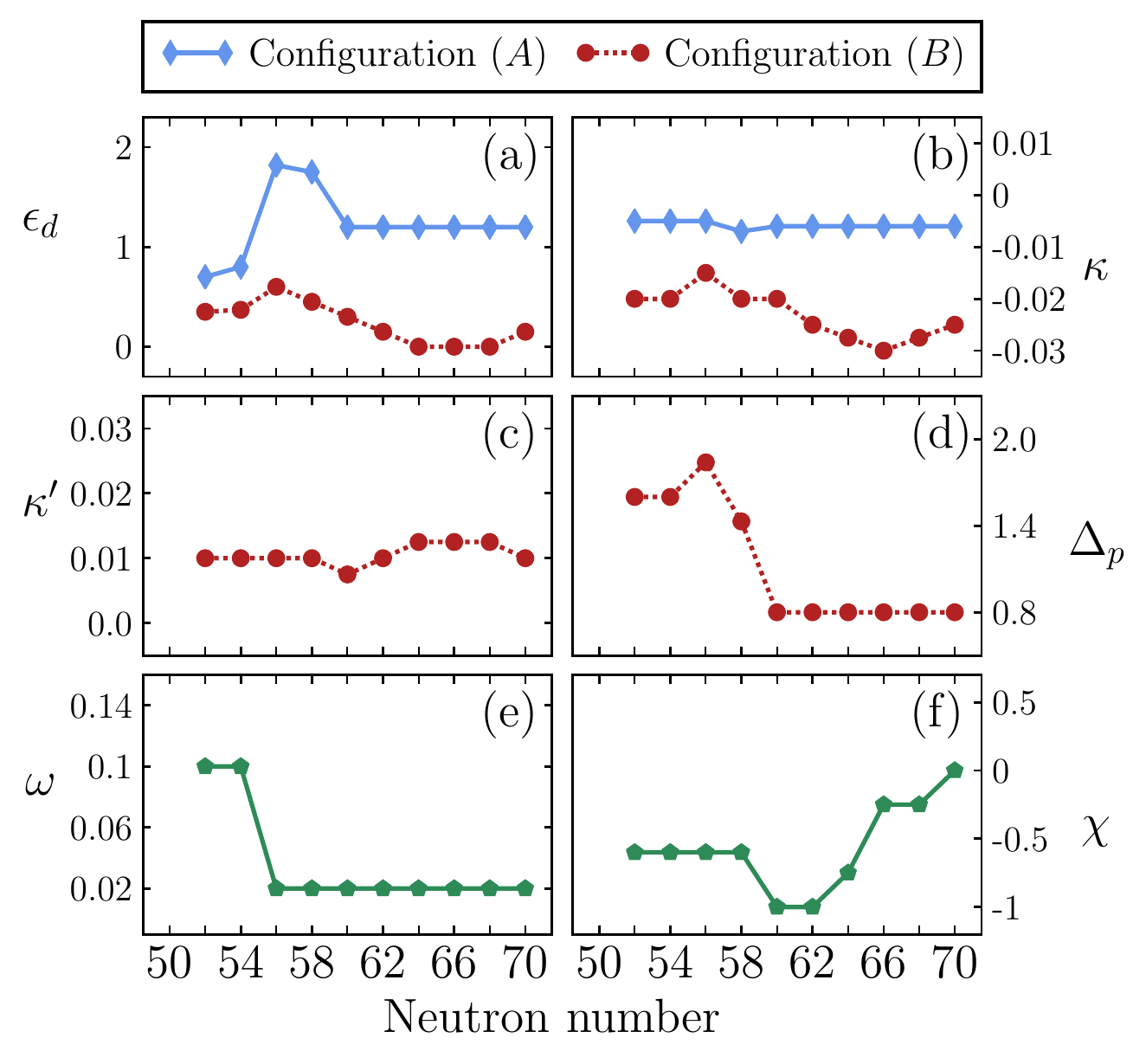}
\caption{
  \small
  Parameters of the IBM-CM Hamiltonians, Eq.~(\ref{Hmat})
  are in MeV and $\chi$ is dimensionless.
  Configurations $A$ and $B$ correspond to boson spaces 
  $[N]$ and $[N\!+\!2]$, respectively, with $N\!=\!1,2,\ldots,8$ ($N\!=\!7,6$),
  for neutron number 52-66 (68,70).
\label{fig:params}}
\end{figure} 

By employing the IBM-CM framework described above,
we have calculated the spectra and other observables
of the entire chain of Zr isotopes, from neutron number 52 to 70. 
The values of the Hamiltonian parameters, obtained by a global fit
to energy and $E2$ data, are shown in Fig.~\ref{fig:params}.
It should be noted that beyond the middle of the 
shell, at neutron number 66, bosons are replaced by boson holes~\cite{ibm}.
Apart from some fluctuations due to 
the subshell closure at neutron number 56
(the filling of the $2d_{5/2}$ orbital), the values of the parameters are 
a smooth function of neutron number and, in some cases, a constant.
A notable exception is the sharp decrease by 1~MeV of
the energy off-set parameter $\Delta_p$
beyond neutron number 56. Such a behavior was observed for the
Mo and Ge chains~\cite{sambataro,duval83} and,
as noted in~\cite{sambataro}, it reflects the effects of the isoscalar
residual interaction between protons and neutrons occupying the
partner orbitals $1g_{9/2}$ and $1g_{7/2}$,
which is the established mechanism for descending cross shell-gap
excitations and onset of deformation in this region~\cite{FedPit79,HeyCas85}.
The $E2$ operator reads 
$\hat{T}(E2)\!=\!e^{(A)}\hat Q^{(N)}_{\chi}+e^{(B)}\hat Q^{(N+2)}_{\chi}$, 
where $\hat{Q}_{\chi}^{(N)}\!=\!\hat{P}_{N}^{\dag }\hat{Q}_{\chi}\hat{P}_{N}$,
$\hat{Q}_{\chi}^{(N+2)}\!=\!P_{N+2}^{\dag }\hat{Q}_{\chi}\hat{P}_{N+2}$
and $\hat{Q}_{\chi}$ is the same operator as in the
Hamiltonian~(\ref{Hmat}).
Here $e^{(A)}\!=\!0.9$ and $e^{(B)}\!=\!2.24$ $({\rm W.u.})^{1/2}$ are the
boson effective charges.
\begin{figure}[t]
\begin{overpic}[width=1\linewidth]{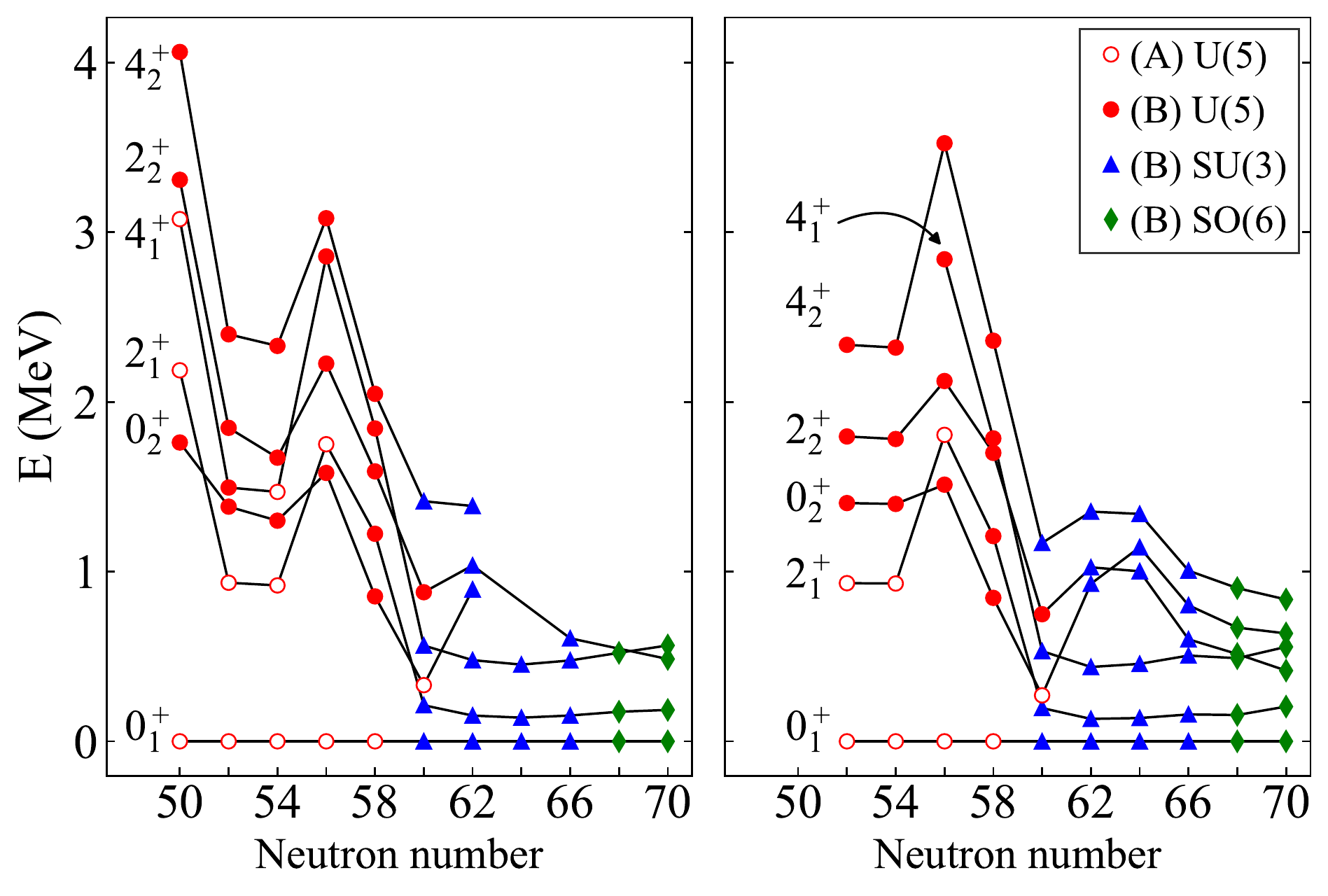}
\put (34,61) {\large (a) {\bf Exp}}
\put (56,61) {\large (b) {\bf Calc}}
\end{overpic}
\caption{
\small
Comparison between (a)~experimental~\cite{Paul17,ensdf} and
(b)~calculated energy levels
$0_{1}^{+},2_{1}^{+},4_{1}^{+},0_{2}^{+},2_{2}^{+},4_{2}^{+}$.
Empty (filled) symbols indicate a state dominated by the 
normal $A$-configuration (intruder $B$-configuration),
with assignments based on the decomposition of Eq.~(\ref{wf}).
The shape of the symbol [$\circ,\,\triangledown,\,\diamond$], 
indicates the closest dynamical symmetry [U(5), SU(3), SO(6)]
relevant to the level considered.
Note that the calculated values
start at neutron number 52, while the experimental values include the
closed shell at 50. Adapted from~\cite{GavLevIac19}.
\label{fig:levels}}
\end{figure}

In Fig.~\ref{fig:levels} we show a comparison between experimental and 
calculated levels. One can see here a rather complex structure.
In the region between neutron number 50 and 56, there appear to be two
configurations, one spherical (seniority-like), ($A$), and one weakly
deformed, ($B$), as evidenced by the ratio $R_{4/2}$, which is at 52-56, 
$R^{(A)}_{4/2}\cong 1.6 $ and  $R^{(B)}_{4/2} \cong 2.3$. 
From neutron number 58, there is a pronounced drop in energy
for the states of configuration~$B$ and at 60, the two configurations
exchange their role indicating a Type~II QPT. 
At this stage,
the intruder configuration~($B$) appears to be at the critical point of
a U(5)-SU(3) Type~I QPT, as evidenced by the low value of the excitation
energy of the first excited $ 0^+ $ state of this configuration
(the $0^{+}_3$ state in $^{100}$Zr shown in Fig.~\ref{fig:spectrum}).
The same situation is seen in the $_{62}$Sm
and $_{64}$Gd isotopes at neutron number 90~\cite{ibm}.
Beyond neutron number 60, the intruder configuration~($B$) is strongly 
deformed, as evidenced by the small value of the
excitation energy of the state $2_{1}^{+}$, $E_{2_{1}^{+}}\!=\!139.3$ keV and by
the ratio $R^{(B)}_{4/2}\!=\!3.24$ in $^{104}$Zr. At still larger
neutron number 66,
the ground state band becomes $\gamma $-unstable
as evidenced by the close energy of the states $2_{2}^{+}$ and $4_{1}^{+}$, 
$E_{2_{2}^{+}}\!=\!607.0$~keV, $E_{4_{1}^{+}}\!=\!476.5$ keV, in $^{106}$Zr, 
and especially by the recent results 
$ E_{4^+_1}\!=\!565$~keV and $ E_{2^+_2}\!=\!485$ keV 
in $^{110} $Zr \cite{Paul17}, a signature of the SO(6) symmetry. 
In this region, the ground state 
configuration undergoes a~crossover from SU(3) to SO(6). 

\begin{figure}[t!]
\centering
\includegraphics[width=1\linewidth]{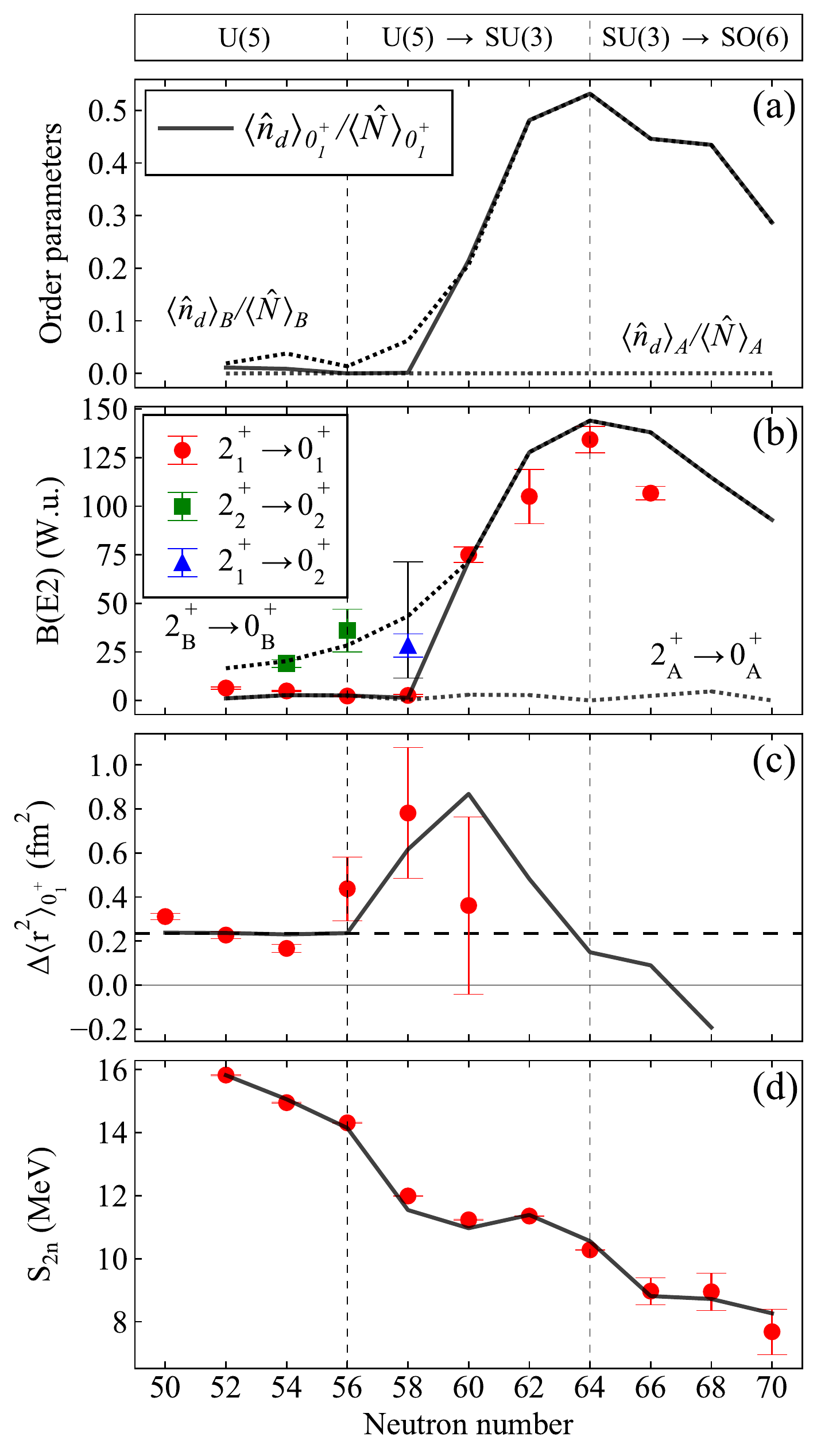}
\caption{
\small
Evolution of order parameters and of observables along the Zr chain.
Symbols (solid lines) denote experimental data (calculated results).
(a)~Order parameters, Eq.~(\ref{order}). Notation of lines is
explained in the text.
(b)~$B(E2)$ values in Weisskopf units (W.u.).
Data taken from~\cite{ensdf,pietralla,Ansari17,Witt18,Singh18}.
Dotted lines denote calculated $E2$ transitions within a configuration.
(c)~Isotope shift, $\Delta\braket{\hat{r}^{2}}_{0^{+}_1}$ 
in fm$^{2}$. Data taken from~\cite{angeli2}. The horizontal 
dashed line at $0.235$ fm$^{2}$ represents the smooth behavior 
in $\Delta \braket{\hat{r}^{2}}_{0^{+}_1}$ due to the $A^{1/3}$ increase 
of the nuclear radius. 
(d)~Two-neutron separation energies, $S_{2n}$, in MeV. 
Data taken from AME2016~\cite{wang-masses}.
Adapted from~\cite{GavLevIac19}.
\label{fig:fig-combined}}
\end{figure}

\begin{figure*}[t]
\centering
\begin{overpic}[width=0.28\linewidth]{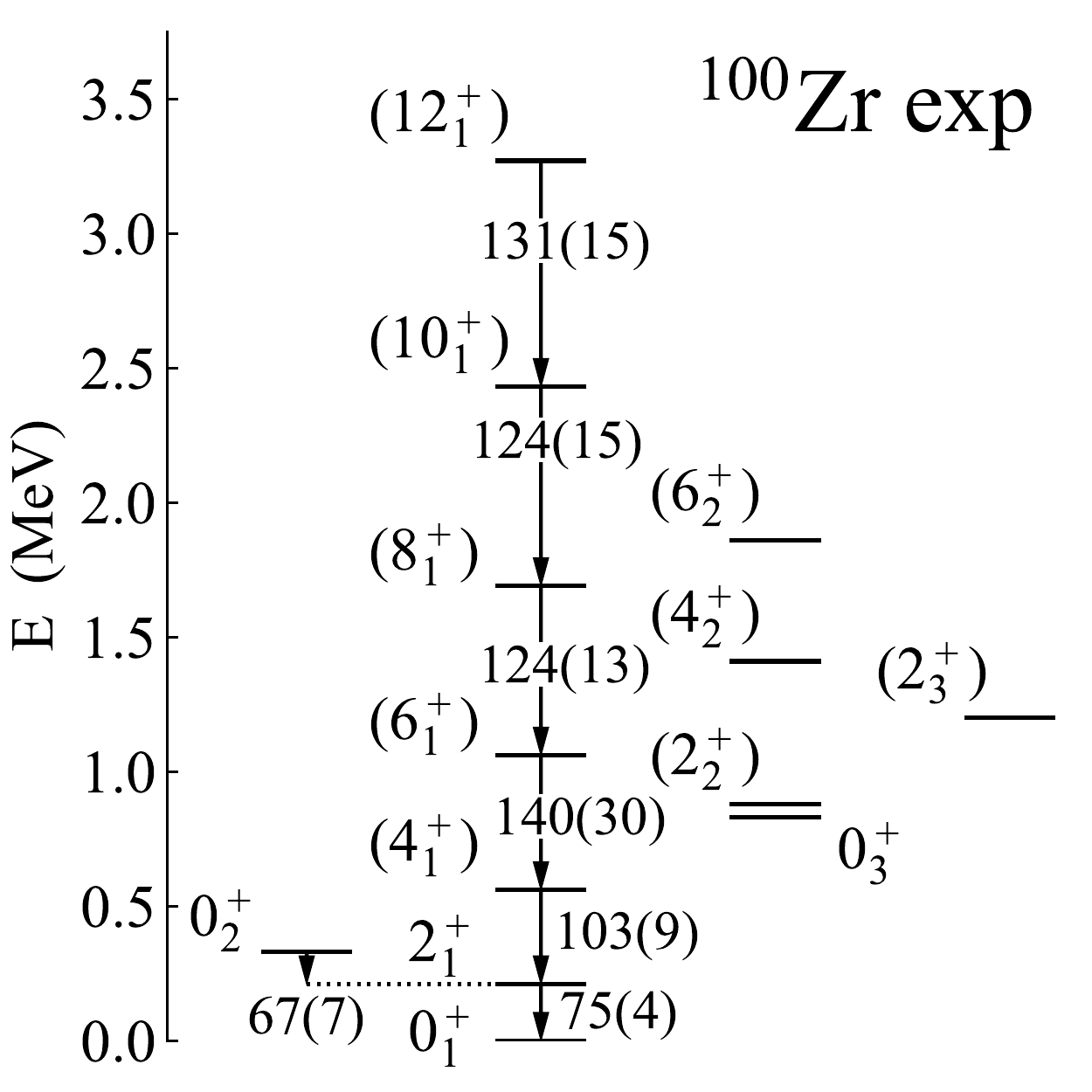}
\put (85,75) {\large(a)}
\end{overpic}
\begin{overpic}[width=0.28\linewidth]{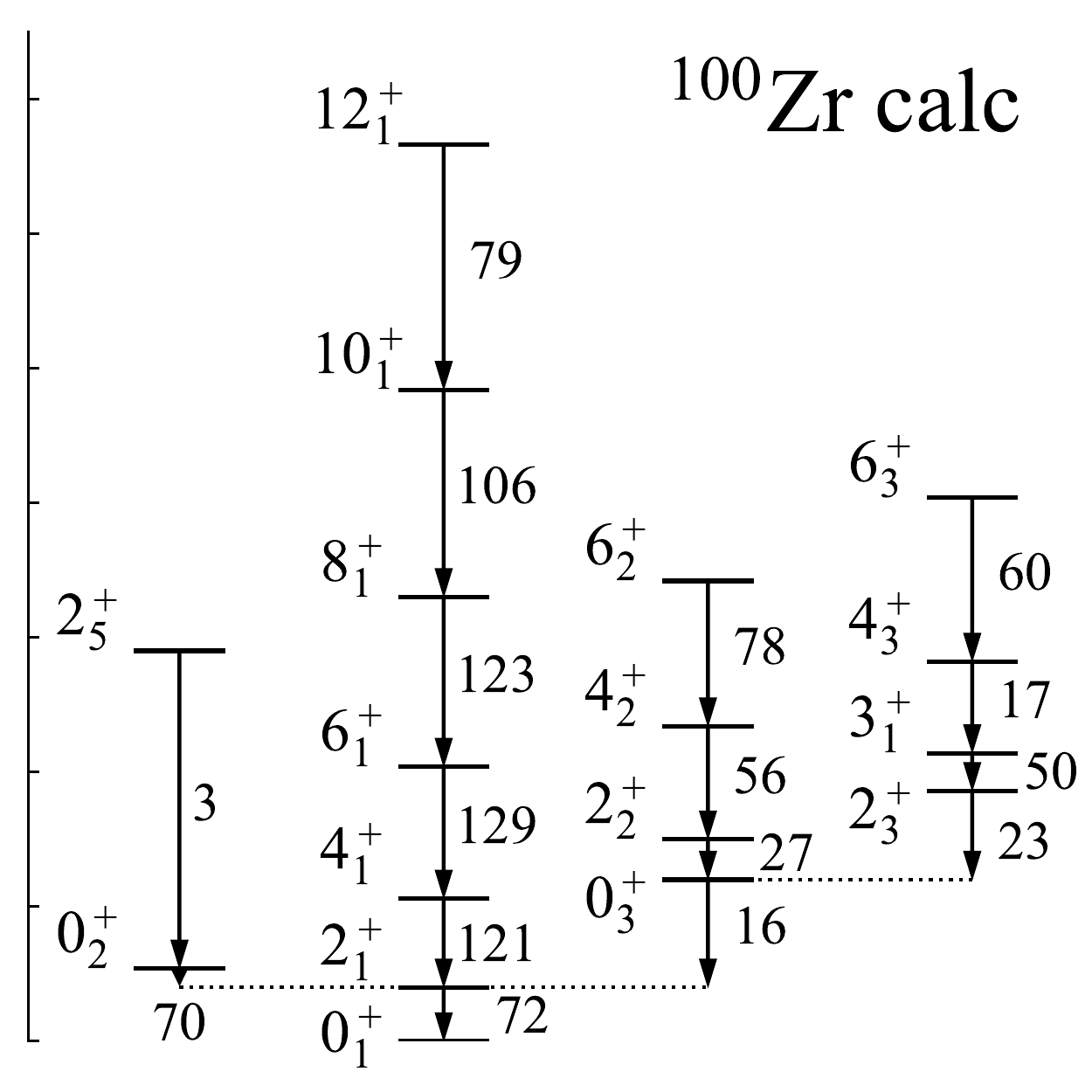}
\put (85,75) {\large(b)}
\end{overpic}
\begin{overpic}[width=0.14\linewidth]{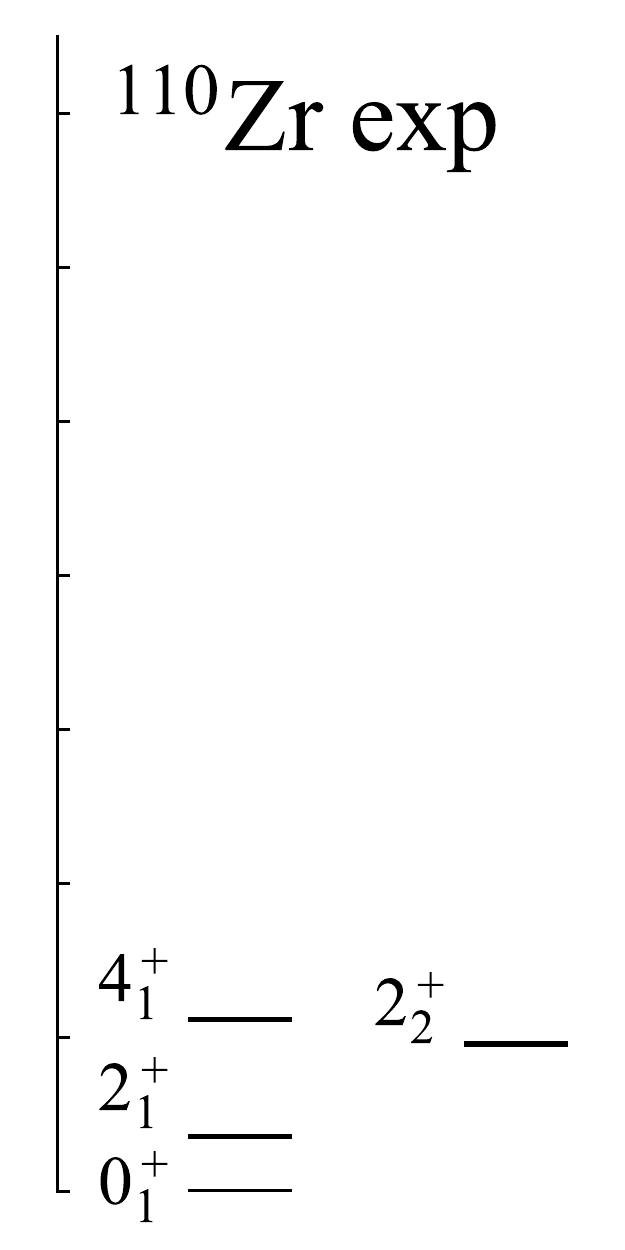}
\put (30,75) {\large(c)}
\end{overpic}
\begin{overpic}[width=0.28\linewidth]{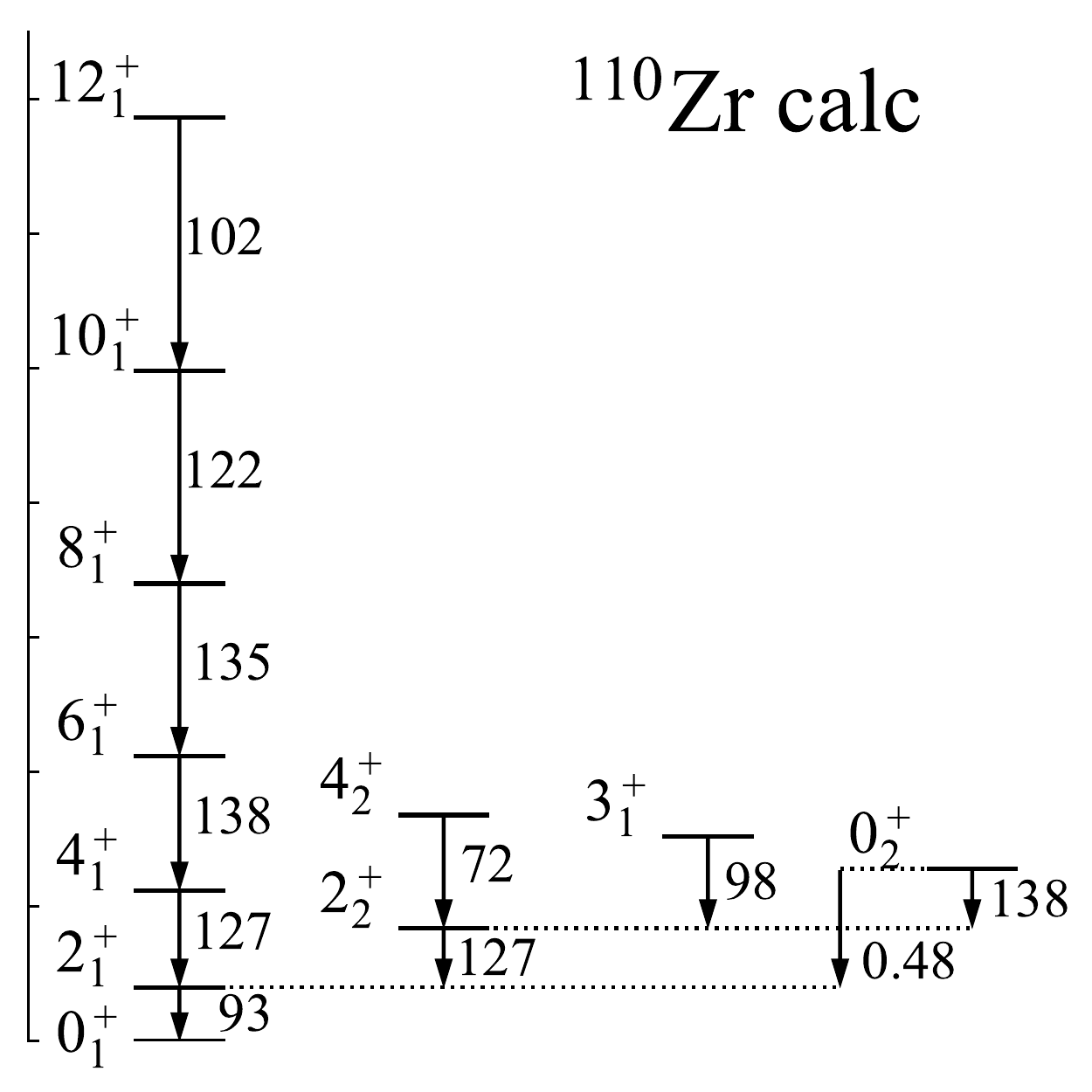}
\put (75,75) {\large(d)}
\end{overpic}
\caption{
\small
Experimental and calculated energy levels in MeV 
and $E2$ rates in W.u. for $^{100}$Zr and $^{110}$Zr.
Adapted from~\cite{GavLevIac19}.
\label{fig:spectrum}}
\end{figure*}

The above spectral analysis signals the presence of coexisting
Type~I and Type~II QPTs, which is the defining property of IQPTs.
In order to understand the nature of these phase transitions,
one needs to study the behavior of the order parameters.
The latter are given by
\ba
\frac{\braket{\hat{n}_d}_A}{\braket{\hat{N}}_A}
\;\;\; , \;\;\;
\frac{\braket{\hat{n}_d}_B}{\braket{\hat{N}}_B}\;\;\; , \;\;\;
\frac{\braket{\hat{n}_d}_{0^{+}_1}}{\braket{\hat{N}}_{0^{+}_1}} ~.
\label{order}
\ea
They involve 
the expectation value of $\hat{n}_d$ in the ground state wave function,
$\ket{\Psi; L\!=\!0^{+}_1}$ and in its
$\Psi_A$ and $\Psi_B$ components~(\ref{wf}),
normalized by the respective boson numbers,
$\braket{\hat{N}}_A\!=\!N$, 
$\braket{\hat{N}}_B\!=\!N\!+\!2$,
$\braket{\hat{N}}_{0^{+}_1}\!=\!a^2N\!+\!b^2(N\!+\!2)$.
Here $\braket{\hat{n}_d}_A$ and $\braket{\hat{n}_d}_B$ portray 
the shape-evolution in configuration~($A$) and ($B$), respectively, 
and $\braket{\hat{n}_d}_{0^{+}_1}= a^2\braket{\hat{n}_d}_A 
+ b^2\braket{\hat{n}_d}_B$
contains information on the normal-intruder mixing.
Fig.~\ref{fig:fig-combined}(a) shows
the evolution of the order parameters involving $\braket{\hat{n}_d}_{A}$
and $\braket{\hat{n}_d}_{B}$ in dotted lines, and $\braket{\hat{n}_d}_{0^{+}_1}$
in solid line. Configuration~($A$) is seen to be spherical for all
neutron numbers considered. 
In contrast, configuration~($B$) is weakly-deformed
for neutron number 52-58. One can see here clearly a jump 
between neutron number 58 and 60 from
configuration~($A$) to configuration~($B$), indicating a 
1$^{st}$ order Type~II phase transition~\cite{frank}, 
a further increase at neutron numbers 60-64
indicating a U(5)-SU(3) Type~I QPT, and, finally,
there is a decrease at neutron number 66, due in part to the 
crossover from SU(3) to SO(6) and in part to the shift from boson 
particles to boson holes after the middle of the major shell 50-82. 
$\braket{\hat{n}_d}_{0^{+}_1}$ is close to $\braket{\hat{n}_d}_A$
for neutron number 52-58 and coincides with $\braket{\hat{n}_d}_B$ 
above 60, indicating a high degree of purity and small
configuration-mixing, with the exception of a narrow transition region.
Indeed, the ground state wave function~(\ref{wf}) has
$a^2=98.2\%,\,b^2=87.2\%$ and $b^2=99.9\%$ for
$^{98}$Zr, $^{100}$Zr and $^{102}$Zr, respectively.
These conclusions are stressed by an analysis of other observables, 
in particular, the $B(E2)$ values.
As shown in Fig.~\ref{fig:fig-combined}(b), the calculated $B(E2)$'s 
agree with the empirical values and follow the same trends as the 
respective order parameters.
\begin{figure*}[]
\centering
\begin{overpic}[width=0.19\linewidth]{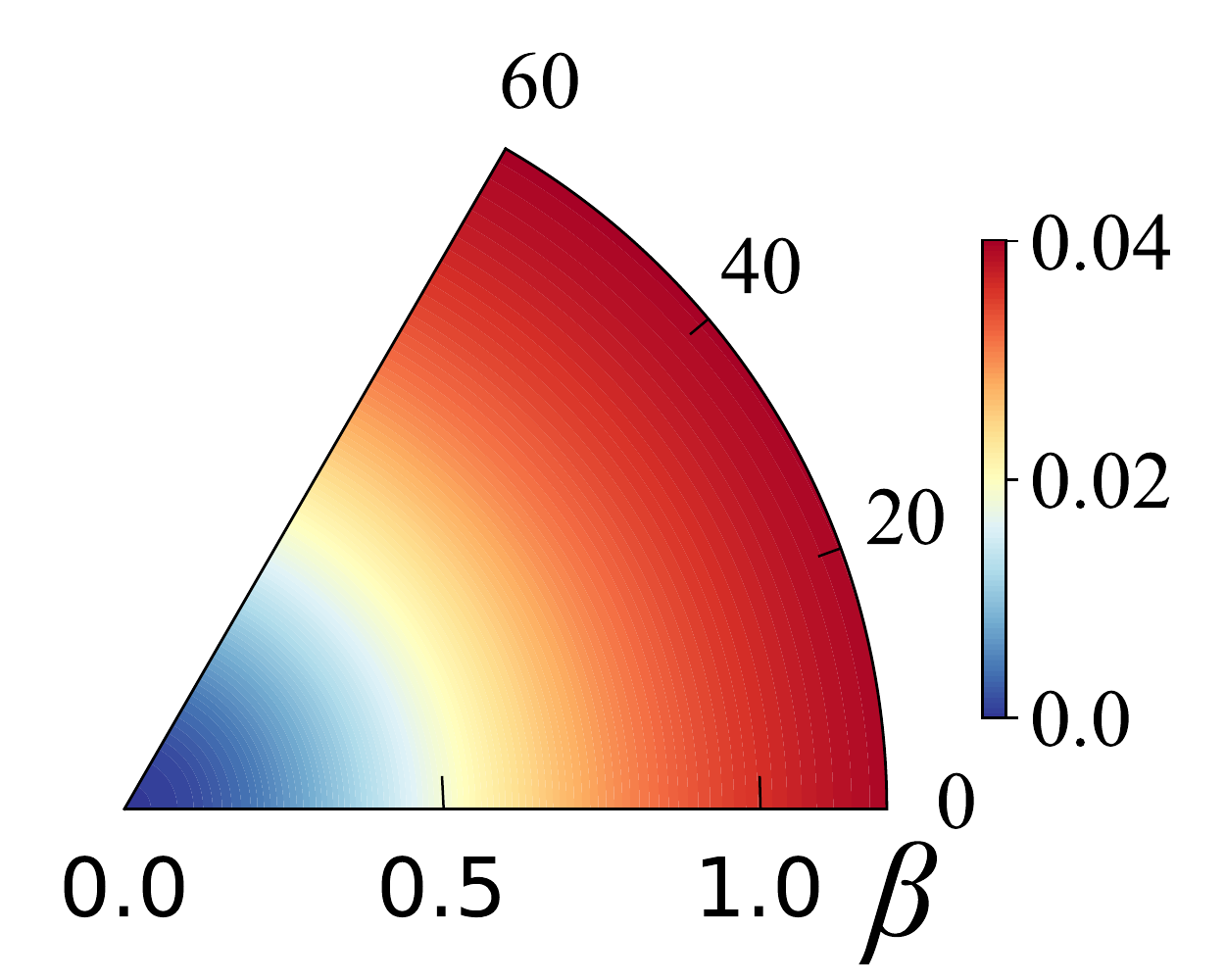}
\put (0,60) {\large $^{92}$Zr}
\end{overpic}
\begin{overpic}[width=0.19\linewidth]{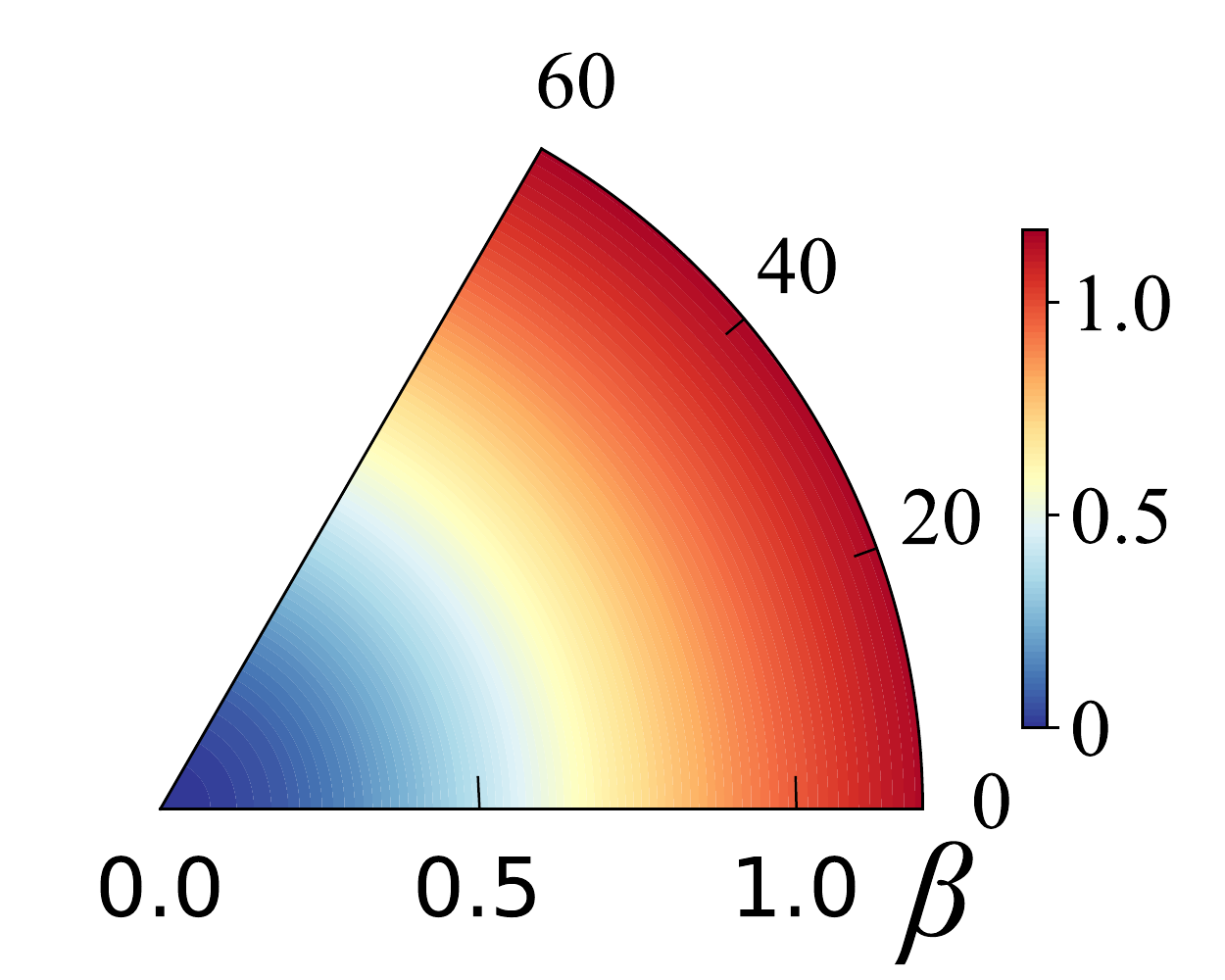}
\put (0,60) {\large $^{94}$Zr}
\end{overpic}
\begin{overpic}[width=0.19\linewidth]{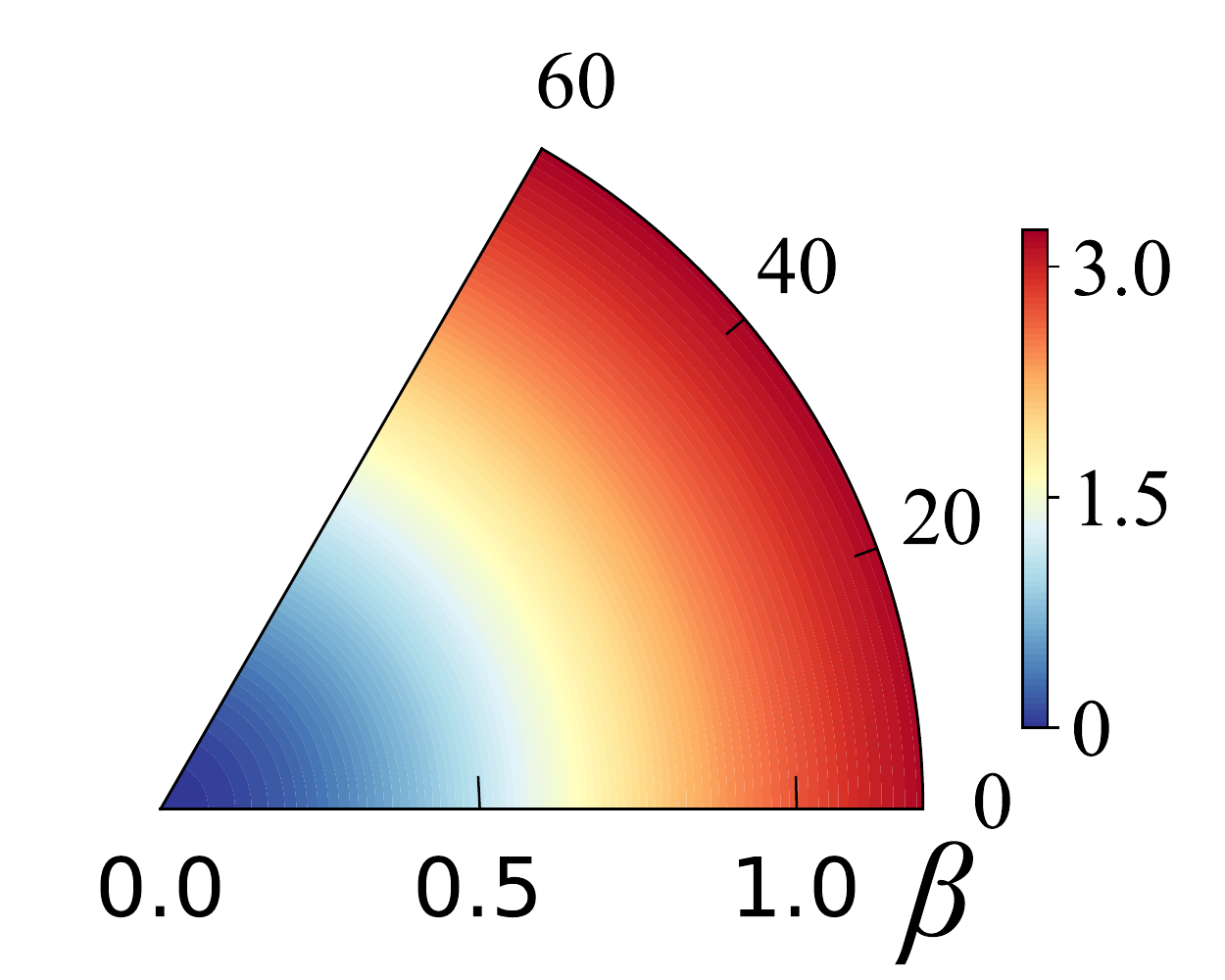}
\put (0,60) {\large $^{96}$Zr}
\end{overpic}
\begin{overpic}[width=0.19\linewidth]{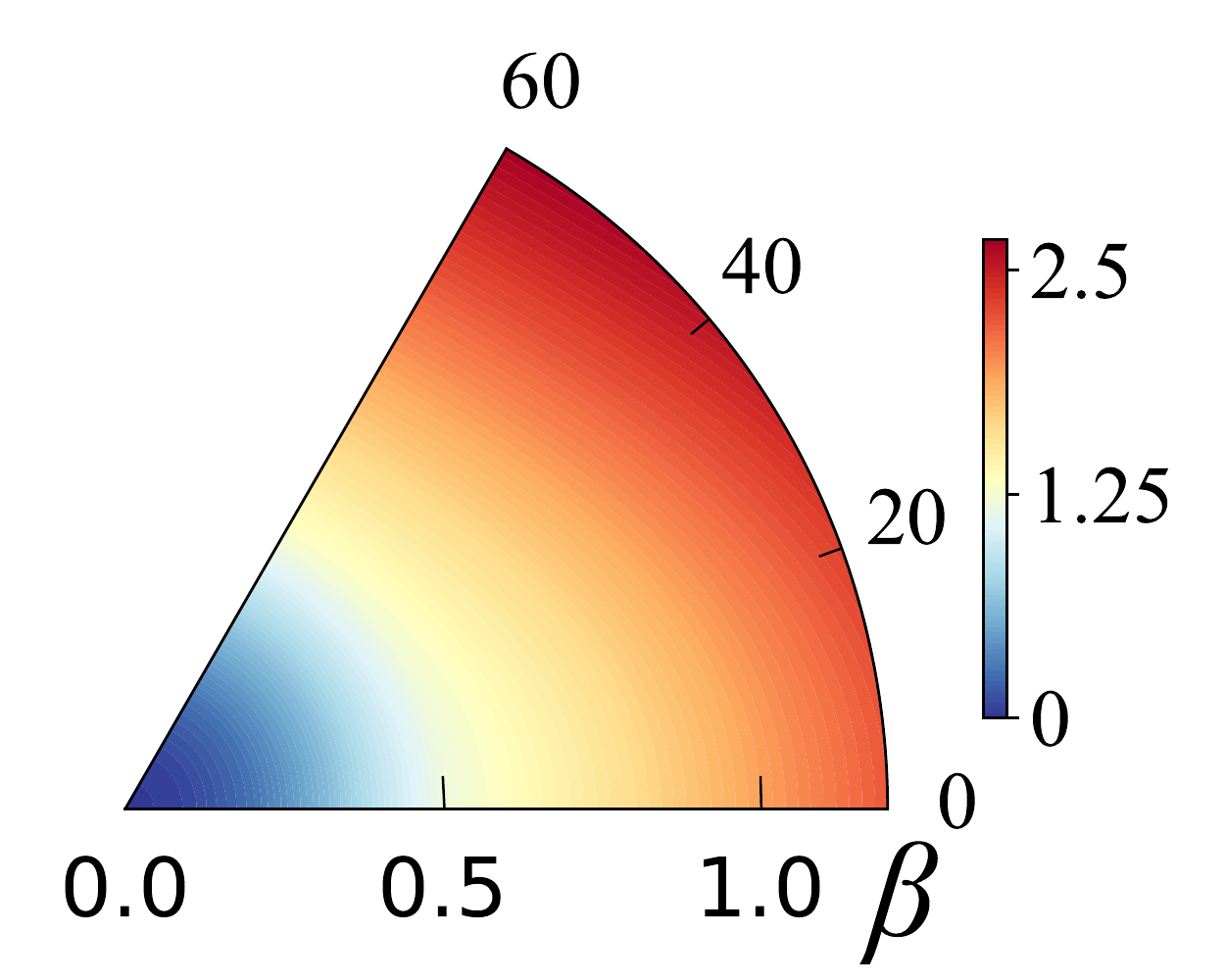}
\put (0,60) {\large $^{98}$Zr}
\end{overpic}
\begin{overpic}[width=0.19\linewidth]{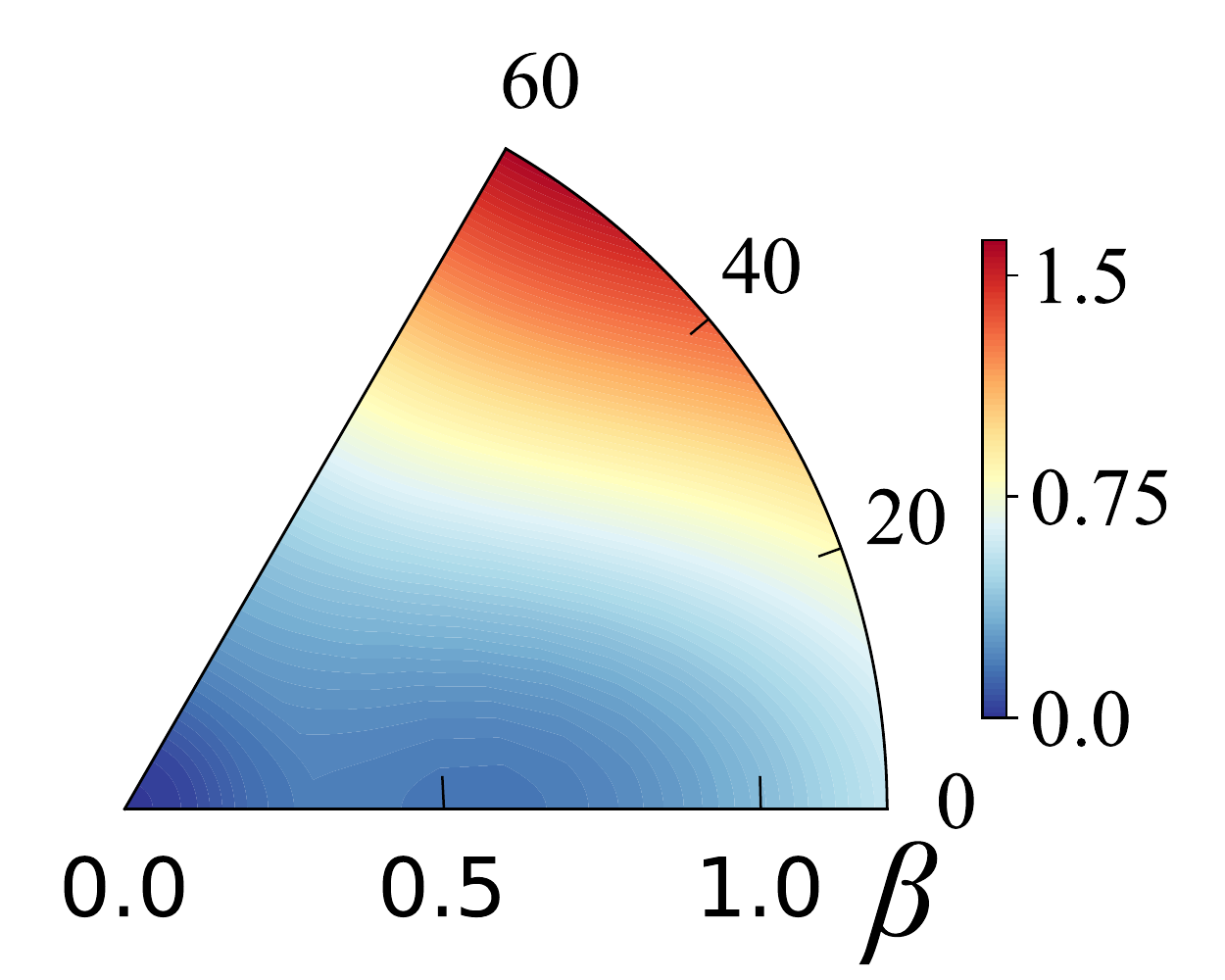}
\put (0,60) {\large $^{100}$Zr}
\end{overpic} \\ 
\begin{overpic}[width=0.19\linewidth]{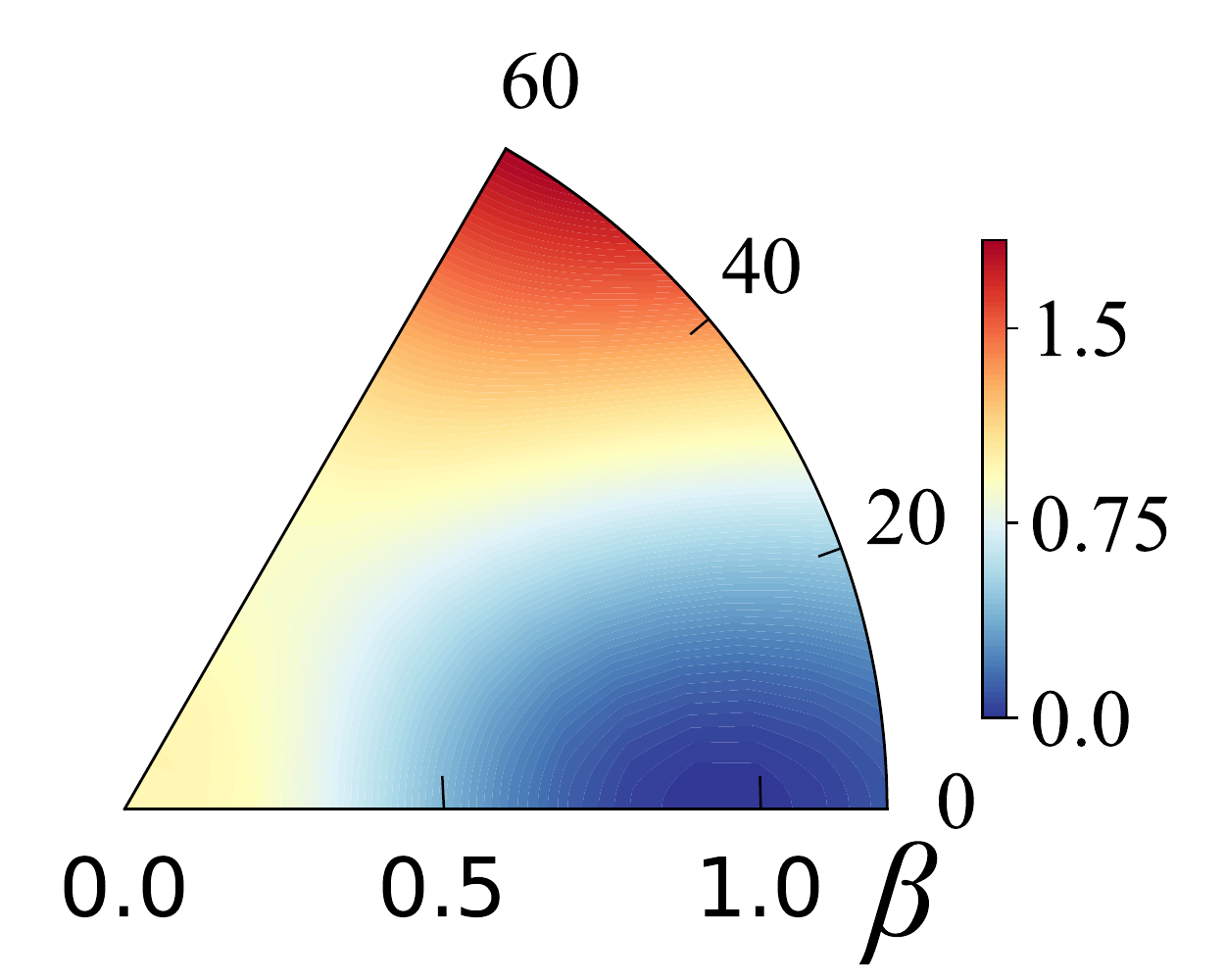}
\put (0,60) {\large $^{102}$Zr}
\end{overpic}
\begin{overpic}[width=0.19\linewidth]{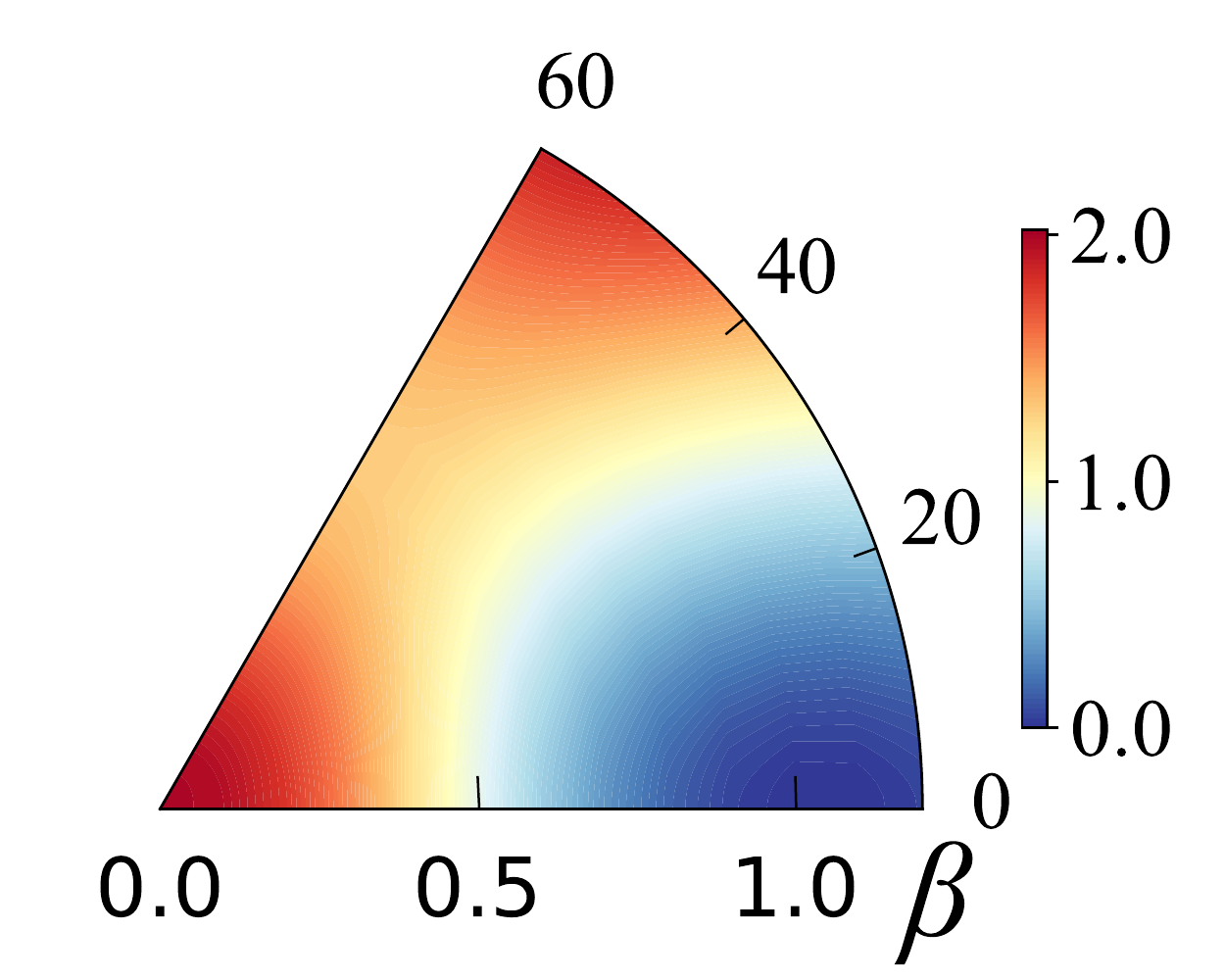}
\put (0,60) {\large $^{104}$Zr}
\end{overpic}
\begin{overpic}[width=0.19\linewidth]{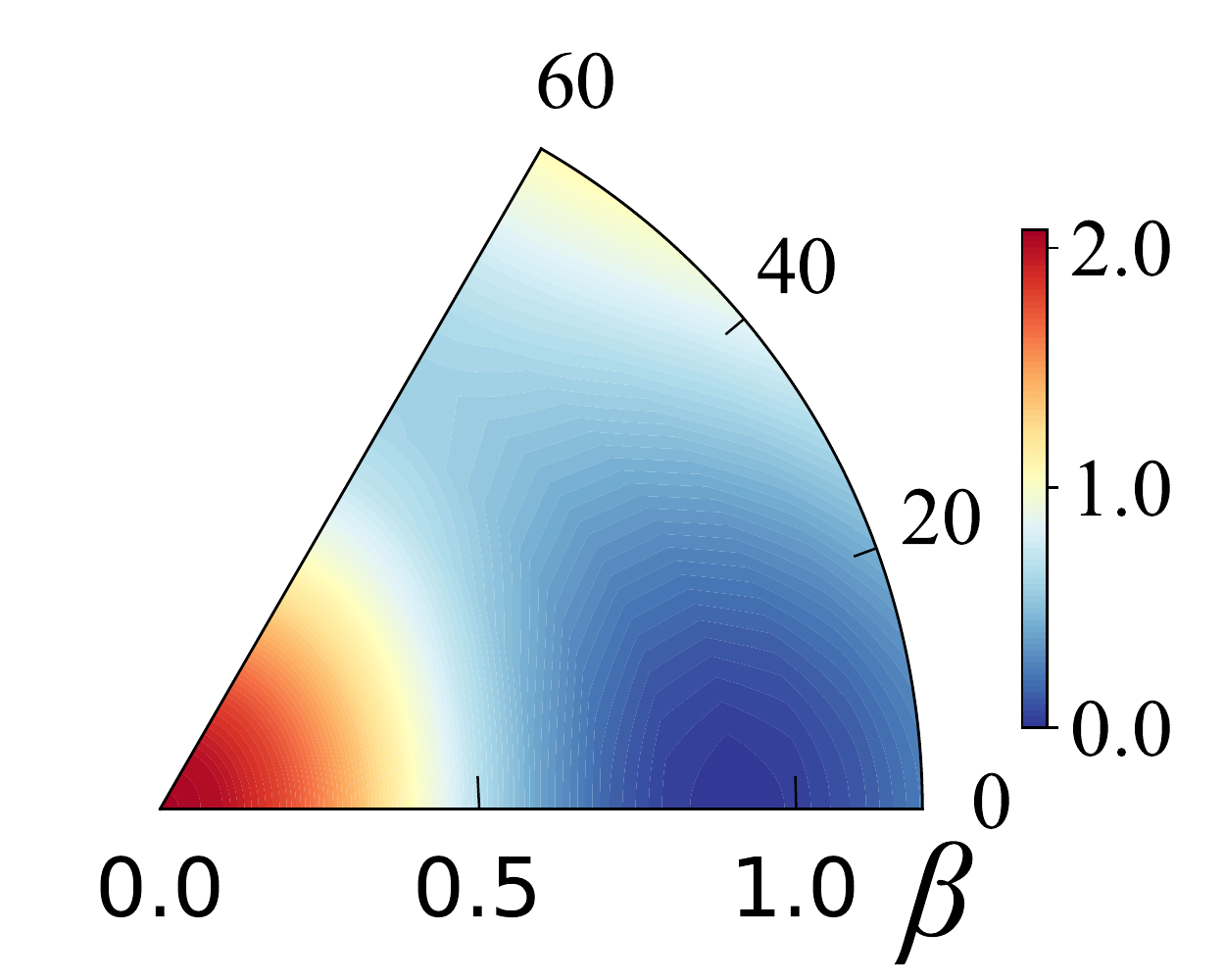}
\put (0,60) {\large $^{106}$Zr}
\end{overpic}
\begin{overpic}[width=0.19\linewidth]{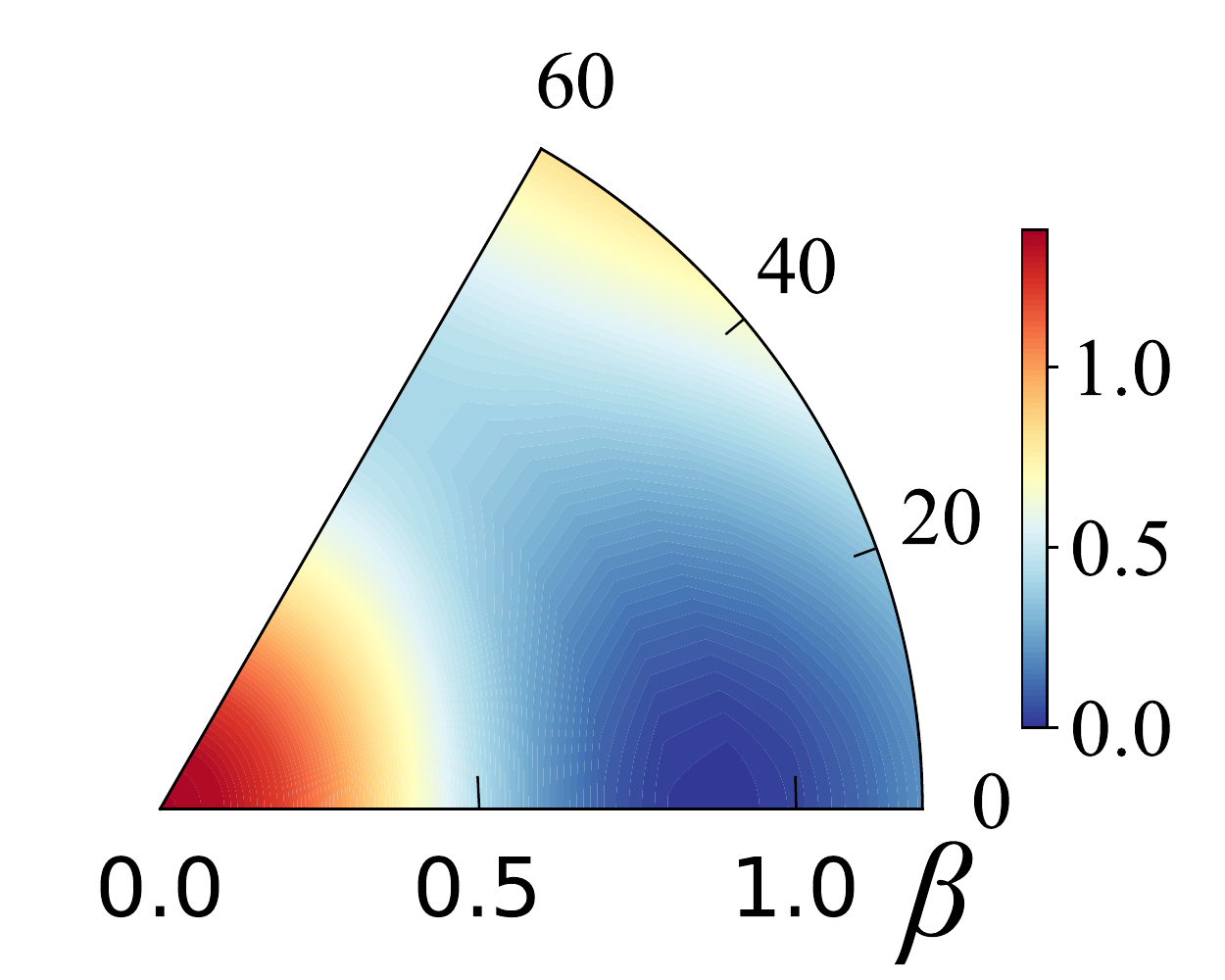}
\put (0,60) {\large $^{108}$Zr}
\end{overpic}
\begin{overpic}[width=0.19\linewidth]{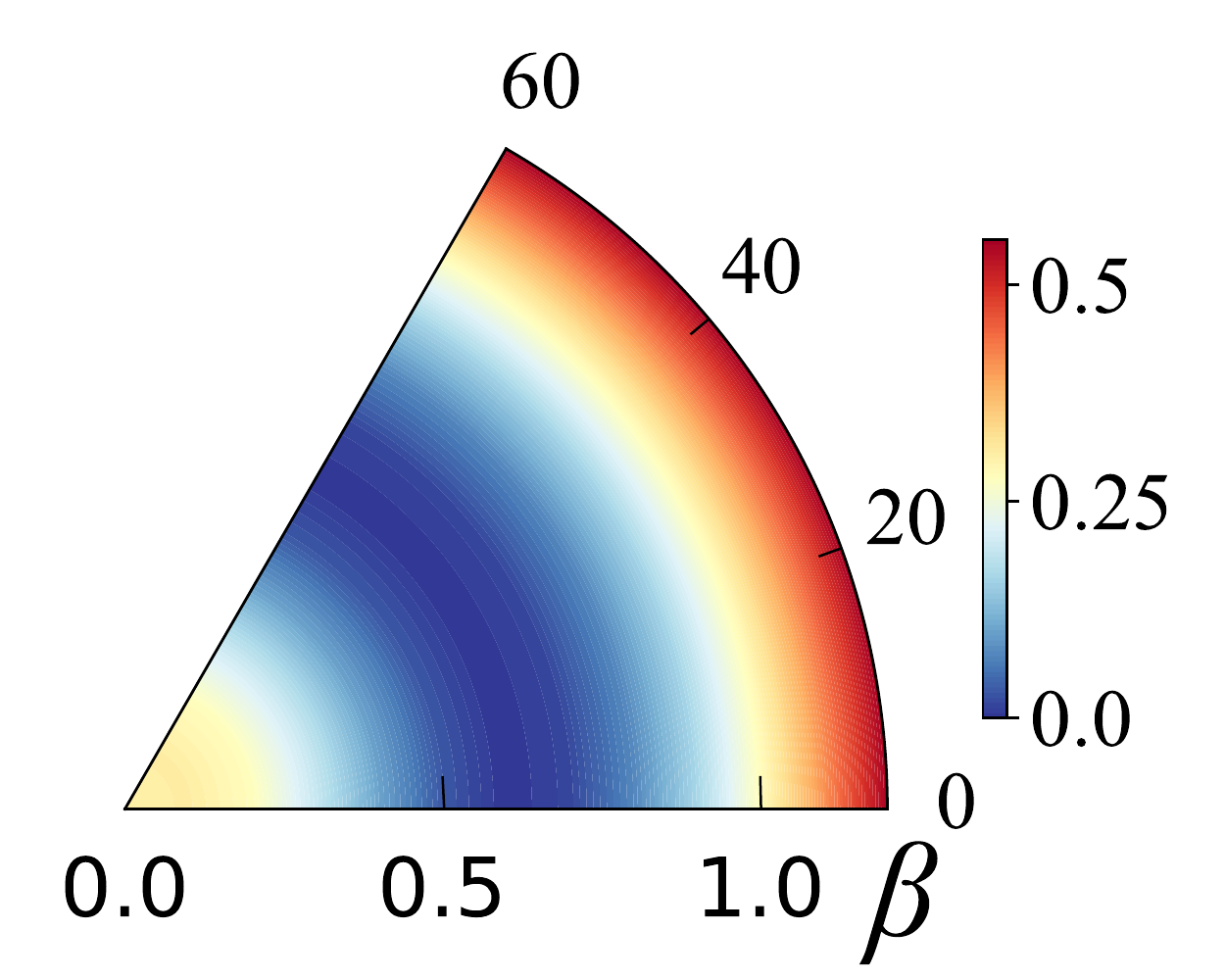}
\put (0,60) {\large $^{110}$Zr}
\end{overpic}
\caption{
\small
Contour plots in the $(\beta ,\gamma )$ plane of the
classical potential surface for the $^{92-110}$Zr isotopes.
Adapted from~\cite{GavLevIac19}.
\label{Eminus}}
\end{figure*}

Further evidence can be obtained from an analysis of the isotope shift 
$\Delta\braket{\hat r^2}_{0^+_1}
=\braket{\hat{r}^{2}}_{0^{+}_1;A+2}-\braket{\hat{r}^2}_{0^{+}_1;A}$, where 
$\braket{\hat r^2}_{0^+_1} $ is the expectation value of $ \hat r^2 $ in the 
ground state $ 0^+_1 $. In the IBM-CM the latter is given by 
$\braket{\hat r^2} = r^2_c + \alpha N_v + \eta [\braket{\hat n_d^{(N)}} 
+ \braket{\hat n_d^{(N+2)}}]$, 
where $r^2_c$ is the square radius of the closed shell, 
$N_v$ is half the number of valence particles, 
and $\eta$ is a coefficient that takes into account the effect 
of deformation~\cite{ibm,vanisacker}.
$\Delta\braket{\hat r^2}_{0^+_1}$ depends on two parameters,
$\alpha\!=\!0.235,\,\eta \!=\! 0.264$ fm$^2$, whose values are fixed
by the procedure of Ref~\cite{vanisacker}.
$\Delta\braket{\hat r^2}_{0^+_1}$ should increase at the transition
point and decrease and, as seen in Fig.~\ref{fig:fig-combined}(c), 
it does so, although the error bars are large and no data are available
beyond neutron number 60.
Similarly, the two-neutron separation energies $S_{2n} $ 
can be written as~\cite{ibm}, 
$S_{2n} = -\tilde{A} -\tilde{B} N_v \pm S^{\text{def}}_{2n} - \Delta_n$, 
where $S^{\text{def}}_{2n}$ is the contribution of the deformation,
obtained by the expectation value of the Hamiltonian in the
ground state~$ 0^+_1$.
The $ + $ sign applies to particles and the $ - $ sign to holes,
and $\Delta_n $ takes into account the neutron subshell closure at 56, 
$\Delta_n = 0 $ for 50-56 and $ \Delta_n = 2 $ MeV for 58-70.
The value of $ \Delta_n $ is taken from Table XII of \cite{barea} 
and $ \tilde{A}\!=\!-16.5,\,\tilde{B}\!=\!0.758$ MeV are determined
by a fit to binding energies of $^{92,94,96}$Zr.
The calculated $ S_{2n}$, shown in Fig. \ref{fig:fig-combined}(d), 
displays a complex behavior. Between neutron number 52 and~56 
it is a straight line, as the ground state is spherical (seniority-like)
configuration~($A$). After 56, it first goes down due to the subshell closure 
at~56, then it flattens as expected from a 1$^{st}$ order Type~I QPT 
(see, for example the same situation in the $_{62}$Sm isotopes~\cite{ibm}).
After 62, it goes down again due to the increasing of deformation and
finally it flattens as expected from a crossover from SU(3) to SO(6).

We note that our calculations describe the experimental data in the
entire range $^{92-110}$Zr very well. Here we show 
only two examples, $^{100}$Zr and $^{110}$Zr. 
$^{100}$Zr is near the critical point of both Type~I and Type~II QPT 
and yet our description of energy levels and $B(E2)$ 
values is excellent, Fig.~\ref{fig:spectrum}(a)-(b). 
The ground state band, configuration~($B$), 
appears to have features of the so-called X(5) 
symmetry~\cite{iachello-x5}, while 
the spherical configuration~($A$) has now become the excited band $ 0^+_2 $. 
$^{110}$Zr, Fig.~\ref{fig:spectrum}(c)-(d),
appears instead to be an excellent example of SO(6) symmetry~\cite{ibm},
although few experimental data are available.
In general, the current results resemble those obtained
in the MCSM~\cite{taka},
however, there are some noticeable differences. Specifically,  
the replacement $\gamma$-unstable $\rightarrow$ triaxial 
and the inclusion of more than two configurations in the MCSM. 
The spherical state in $^{100}$Zr is identified in the MCSM as $0^{+}_4$,
in contrast to $0^{+}_2$ in the current calculation and the data.
Both calculations show a large jump in 
$B(E2;2^+_1\rightarrow0^+_1)$, between $ ^{98} $Zr and $ ^{100} $Zr, 
typical of a 1$^{st} $ order QPT. This is in contrast with mean-field 
based calculations~\cite{delaroche,mei,nomura16}, 
which due to their character smooth out the phase transitional behavior, 
and show no such jump at the critical point of the QPT
(see Fig.~2 of~\cite{Singh18}).
The observed peak in $B(E2;2^+_1\rightarrow0^+_1)$ for $^{104}$Zr,
is reproduced here but not by the MCSM.

The algebraic approach allows both
a quantum and a classical analysis of QPTs.
Classical potential surfaces are obtained by the method
of matrix-coherent-states~\cite{frank}. 
As seen in Fig.~\ref{Eminus}, the calculated surfaces
confirm the quantum results, as they show a transition from spherical 
($^{92-98}$Zr), to a flat-bottomed potential at $^{100}$Zr,
to axially deformed ($^{102-104}$Zr), and finally to 
$\gamma $-unstable ($^{106-110}$Zr).

This work was supported in part by U.S. DOE under Grant No.
DE-FG02-91ER-40608 and by the US-Israel Binational Science Foundation 
Grant No. 2016032.


\begin{thebibliography}{}

\bibitem{cejnar} 
P. Cejnar, J. Jolie and R.~F. Casten, 
Rev. Mod. Phys. \textbf{82}, 2155 (2010).

\bibitem{Heyde11}
K.~Heyde and J.~L.~Wood,
Rev. Mod. Phys. \textbf{83}, 1467 (2011).

\bibitem{frank} 
A. Frank, P. Van Isacker and F. Iachello, 
Phys. Rev. C \textbf{73}, 061302(R) (2006).

\bibitem{ramos14}
J.~E.~Garc\'\i a-Ramos and K.~Heyde,
Phys. Rev. C \textbf{89}, 014306 (2014);
Phys. Rev. C \textbf{92}, 034309 (2015).

\bibitem{GavLevIac19}
N. Gavrielov, A. Leviatan and F. Iachello,
Phys. Rev. C \textbf{99}, 064324 (2019).

\bibitem{pietralla} 
C. Kremer \textit{et al.},
Phys. Rev. Lett. \textbf{117}, 172503 (2016).

\bibitem{Ansari17}
S. Ansari \textit{et al.},
Phys. Rev. C \textbf{96}, 054323 (2017).

\bibitem{Paul17}
N. Paul \textit{et al.},
Phys. Rev. Lett. \textbf{118}, 032501 (2017).

\bibitem{Witt18}
W. Witt \textit{et al.},
Phys. Rev. C \textbf{98}, 041302(R) (2018).

\bibitem{Singh18}
P. Singh \textit{et al.},
Phys. Rev. Lett. \textbf{121}, 192501 (2018).

\bibitem{delaroche}
J.-P. Delaroche \textit{et al.},
Phys. Rev. C \textbf{81}, 014303 (2010).

\bibitem{mei}
H. Mei {\it et al.},
Phys. Rev. C \textbf{85}, 034321 (2012).

\bibitem{nomura16}
K. Nomura {\it et al.},
Phys. Rev. C \textbf{94}, 044314 (2016).

\bibitem{taka} 
T. Togashi, Y. Tsunoda, T. Otsuka and N. Shimizu,
Phys. Rev. Lett. \textbf{117}, 172502 (2016).

\bibitem{ibm} 
F. Iachello and A. Arima, 
\textit{The Interacting Boson Model}, 
(Cambridge Univ. Press, Cambridge, 1987).

\bibitem{duval}
P.~D. Duval and B.~R. Barrett, 
Phys. Lett. B \textbf{100}, 223 (1981); 
Nucl. Phys. A \textbf{376}, 213 (1982).

\bibitem{sambataro} 
M. Sambataro and G. Moln\'ar, 
Nucl. Phys. A \textbf{376}, 201 (1982).

\bibitem{duval83}
P.~D. Duval \textit{et al.},
Phys. Lett. B \textbf{124}, 297 (1983).
  
\bibitem{FedPit79}
P. Federman and S. Pittel,
Phys. Rev. C \textbf{20}, 820 (1979).

\bibitem{HeyCas85}
K Heyde \textit{et al.},
Phys. Lett. B \textbf{155}, 303 (1985).

\bibitem{ensdf}
Evaluated Nuclear Structure Data File (ENSDF),
\href{https://www.nndc.bnl.gov/ensdf/}{https://www.nndc.bnl.gov/ensdf/}.

\bibitem{angeli2} 
I. Angeli and K.~P. Marinova, 
At. Data Nucl. Data Tables \textbf{99}, 69 (2013).

\bibitem{wang-masses} 
M. Wang \textit{et al.},
Chinese Phys. C \textbf{41}, 030003 (2017).

\bibitem{vanisacker} 
S. Zerguine \textit{et al.},
Phys. Rev. C \textbf{85}, 034331 (2012).

\bibitem{barea}
J. Barea and F. Iachello, 
Phys. Rev. C \textbf{79}, 044301 (2009).

\bibitem{iachello-x5}
F. Iachello,
Phys. Rev. Lett. \textbf{87}, 052502 (2001).

\end{thebibliography}
\end{document}